\documentclass[12pt]{article}
\usepackage{a4wide,amsmath,amssymb,alltt,feynarts,graphicx,axodraw,colordvi}

\newcommand\ie{i.e.\ }
\newcommand\eg{e.g.\ }
\newcommand\backsl{\symbol{92}}
\newcommand\lbrac{\symbol{123}}
\newcommand\rbrac{\symbol{125}}

\newcommand\Brac[1]{\lbrac#1\rbrac}
\newcommand\Code[1]{\ensuremath{\texttt{#1}}}
\newcommand\Var[1]{\ensuremath{\mathit{#1}}}
\newcommand\icon[1]{\lower 6pt\hbox{\includegraphics[scale=.8]{#124}}}

\makeatletter
\def\reportno#1{\gdef\@reportno{#1}}
\def\@maketitle{%
  \hfill{\small\begin{tabular}[t]{r}%
    \@reportno
  \end{tabular}\par}%
  \vskip 2em%
  \begin{center}%
    \let \footnote \thanks
    {\large \@title \par}%
    \vskip 1.5em%
    {
      \lineskip .5em%
      \begin{tabular}[t]{c}%
        \@author  
      \end{tabular}\par}%
    \vskip 1em%
    {
     \@date}%
  \end{center}%
  \par
  \vskip 1.5em}
\makeatother

\begin{document}

\reportno{MPP--2007--166\\hep--ph/yymm.nnnn}

\title{FeynEdit -- a tool for drawing Feynman diagrams}

\author{T. Hahn$^a$, P. Lang$^b$ \\
$^a$Max-Planck-Institut f\"ur Physik \\
F\"ohringer Ring 6, D--80805 Munich, Germany \\
$^b$Rieslingstr.\ 62, D--74343 Lauffen}

\date{November 8, 2007}

\maketitle

\begin{abstract}
We describe the FeynEdit tool for drawing Feynman diagrams.  Input and
output is done using the \LaTeX\ macros of FeynArts, which also implies
that diagrams drawn by FeynArts can be edited with FeynEdit.  The 
\LaTeX\ code can be conveniently transferred using copy-and-paste.
\end{abstract}


\section{Introduction}

The FeynArts package \cite{FA} can paint Feynman diagrams and export 
them as \LaTeX\ code, such that they can be included directly in 
publications.  For example, the diagram
\begin{center}
\vspace*{-6ex}
\begin{feynartspicture}(150,150)(1,1)
   \FADiagram{}
   \FAProp(0.,10.)(6.,10.)(0.,){/Straight}{0}
   \FALabel(3.,9.18)[t]{$1$}
   \FAProp(20.,10.)(14.,10.)(0.,){/Straight}{0}
   \FALabel(17.,10.82)[b]{$2$}
   \FAProp(6.,10.)(14.,10.)(0.8,){/Straight}{0}
   \FALabel(10.,5.98)[t]{$3$}
   \FAProp(6.,10.)(14.,10.)(-0.8,){/Straight}{0}
   \FALabel(10.,14.02)[b]{$4$}
   \FAVert(6.,10.){0}
   \FAVert(14.,10.){0}
\end{feynartspicture}
\vspace*{-6ex}
\end{center}
is represented by the \LaTeX\ code
\begin{verbatim}
   \FAProp(0.,10.)(6.,10.)(0.,){/Straight}{0}
   \FALabel(3.,9.18)[t]{$1$}
   \FAProp(20.,10.)(14.,10.)(0.,){/Straight}{0}
   \FALabel(17.,10.82)[b]{$2$}
   \FAProp(6.,10.)(14.,10.)(0.8,){/Straight}{0}
   \FALabel(10.,5.98)[t]{$3$}
   \FAProp(6.,10.)(14.,10.)(-0.8,){/Straight}{0}
   \FALabel(10.,14.02)[b]{$4$}
   \FAVert(6.,10.){0}
   \FAVert(14.,10.){0}
\end{verbatim}
The elements of the diagram are easy to recognize and it is
straightforward to make changes \eg to the label text.  It is less
straightforward, however, to alter the geometry of the diagram, \ie to
move vertices and propagators.

The new tool FeynEdit fills this gap by allowing the user to 
copy-and-paste their \LaTeX\ code of the Feynman diagram into the editor, 
visualize the diagram, modify it using the mouse, and finally 
copy-and-paste it back into the text.


\section{Installation}

The FeynEdit package can be downloaded from
\Code{http://www.feynarts.de}.  Unpack the tar-file and run \Code{make},
for example:
\begin{alltt}
   gunzip -c FeynEdit-\(n\).\(m\).tar.gz | tar xvf -
   cd FeynEdit-\(n\).\(m\)
   make
\end{alltt}
The package contains both the source files and the compiled Java
byte-code.  The actual Java program is \Code{FeynEdit.jar} and the above
\Code{make} command only turns the jar-file into a regular executable by
adding the script signature with the full path of the local system's
Java interpreter.  On Windows with a Java Runtime installed
(\Code{http://www.java.com}), \Code{FeynEdit.jar} executes directly when
double-clicked.


\section{Usage}

\begin{figure}
\centerline{\includegraphics{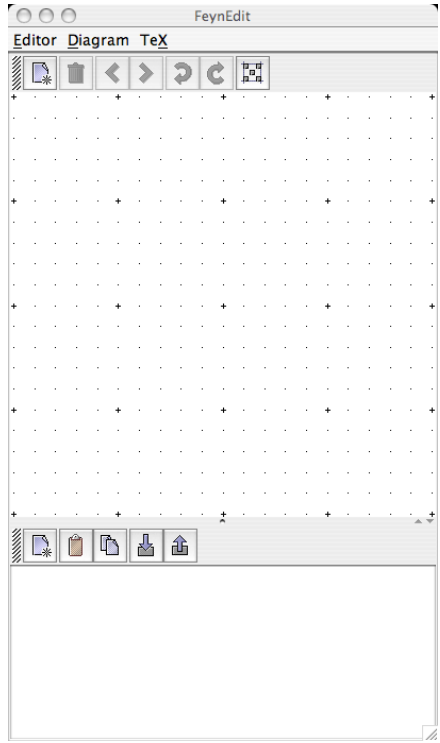}}
\caption{\label{fig:startup}The start-up screen of FeynEdit.}
\end{figure}

The editor is started by typing \Code{FeynEdit} at the command line or
by double-clicking on the \Code{FeynEdit.jar} icon on Windows.  The
start-up screen is shown in Fig.~\ref{fig:startup}.

The window is divided into an upper panel for the diagram display and a
lower panel which shows the \LaTeX\ code.  To display an existing
Feynman diagram, mark its \LaTeX\ code with the mouse and paste it into
the lower dialog box.  Then press the \icon{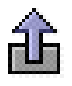}\,button to display
the diagram.  Otherwise, start with an empty canvas and use the mouse to
add elements.

When finished with editing, press the \icon{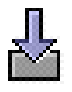}\,button to turn the
diagrams into \LaTeX\ code, then pick up the latter with the mouse and
paste it (back) into your text.

\begin{figure}
\centerline{\includegraphics{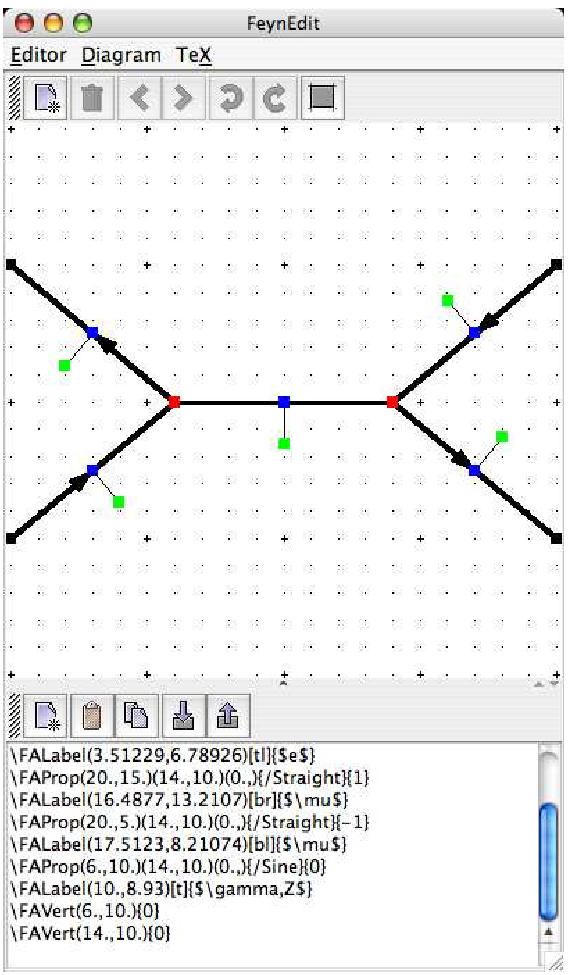}\qquad\includegraphics{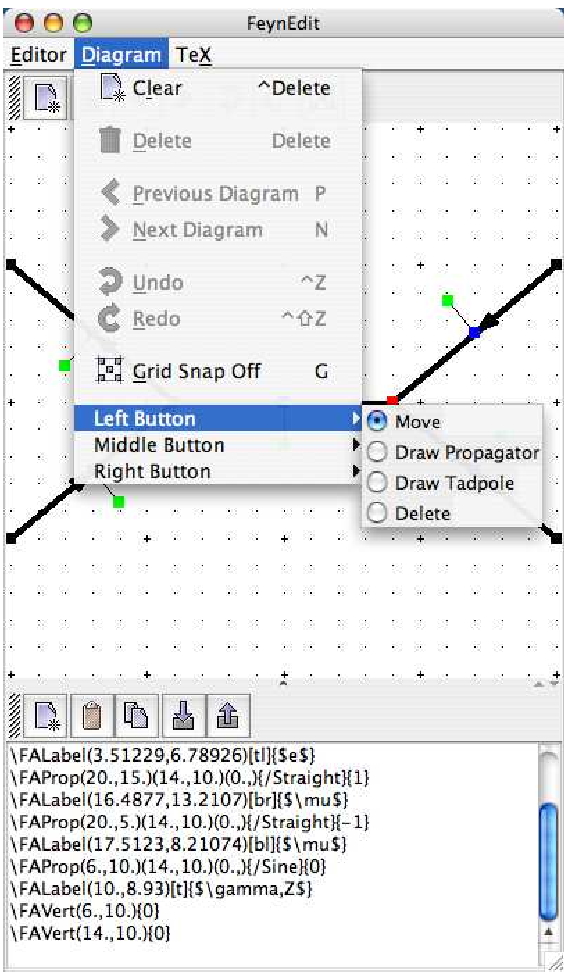}}
\caption{\label{fig:diagram}Left panel: The diagram pasted into FeynEdit 
and displayed.  Right panel: The Mouse Button Assignment Menu in FeynEdit.}
\end{figure}

Just as in the Topology Editor of FeynArts, vertices are marked with
red, propagators with blue, and labels with green squares.  When 
clicking on a square, the corresponding entity becomes marked and the 
square is drawn a little larger.

Propagators come in two varieties which have to be distinguished for
editing purposes: tadpoles, with coincident initial and final vertices,
and `ordinary' propagators.

The diagram can be edited with the mouse.  Four editing functions are
available:
\begin{itemize}
\item Move vertices, propagators, and labels:
      Click on the corresponding box (red, blue, green) and drag
      it to the desired position.

\item Draw tadpoles:
      Click on the `footpoint' of the tadpole and drag it to the
      desired size and orientation.

\item Draw `ordinary' propagators:
      Click on the starting point and drag to the end point.

\item Delete objects:
      Click on the square (red, blue, green) corresponding to the
      object you want to delete.  When deleting a vertex, the 
      propagators adjacent to this vertex are also deleted.  When 
      deleting a propagator, the corresponding label is also deleted.
\end{itemize}
In the default setup, the left mouse button moves objects, the middle 
mouse button draws tadpoles, and the right mouse button draws 
propagators.  The assignment of the mouse button can be changed in the 
Mouse Button Assignment menu (Fig.~\ref{fig:diagram}, right panel).

Following is an overview of all buttons:
\begin{itemize}
\item[\icon{Export}] Interpret the \LaTeX\ code and display the 
diagrams.

\item[\icon{Import}] Turn the edited diagrams into \LaTeX\ code.

\item[\icon{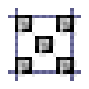}] Turn on Grid Snap, \ie restrict a dragged
object's location to lie on a grid position (`quantize' the drag
movement).  This is to aid the aligned placement of items.

\item[\icon{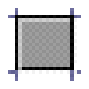}] Turn off Grid Snap, \ie allow objects to be
dragged to arbitrary positions.

\item[\icon{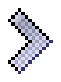}] Move forward one diagram.  For this button to 
become active, the \LaTeX\ code must contain more than one diagram, 
separated by \Code{\backsl FADiagram} directives.

\item[\icon{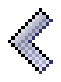}] Move back one diagram.

\item[\icon{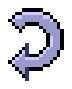}] Undo last change.

\item[\icon{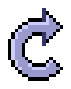}] Redo last undone change.

\item[\icon{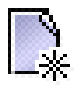}] Clear panel.

\item[\icon{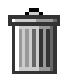}] Delete the currently marked entity.

\item[\icon{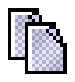}] Copy the contents of the \LaTeX\ panel into the 
copy-and-paste buffer.

\item[\icon{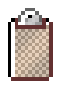}] Paste the current copy-and-paste buffer into the 
\LaTeX\ panel.
\end{itemize}

Note that details of the diagram, such as line attributes and label
texts, are neither displayed by the editor, nor can they be specified
for new tadpoles and propagators.  Thus, for example, a gluon line is
not drawn as a cycloid.  This is largely for performance reasons (think
of dragging a gluon line) and may be added in a future version.  At any
rate, it is not a serious deficit because that information can easily be
added in the \LaTeX\ code.  For instance, the line
\begin{verbatim}
   \FAProp(0.,10.)(6.,10.)(0.,){/Straight}{0}
\end{verbatim}
refers to a straight line.  Simply changing ``\Code{Straight}'' to 
``\Code{Cycles}'' makes it a gluon line.  The next section gives all 
such details on the FeynArts \LaTeX\ style.

\section{Graphics Primitives in feynarts.sty}

The FeynArts style is included in a \LaTeX\ $2\varepsilon$ document with
\begin{verbatim}
  \usepackage{feynarts}
\end{verbatim}
It makes three graphics primitives available with which Feynman diagrams
can be drawn:
\begin{itemize}
\item \Code{\backsl FAProp} draws a propagator,
\item \Code{\backsl FAVert} draws a vertex,
\item \Code{\backsl FALabel} places a label.
\end{itemize}
In addition, it provides formatting/geometry directives:
\begin{itemize}
\item \Code{\backsl begin}$\dots$\Code{end\Brac{feynartspicture}}
  delineates a sheet of Feynman diagrams,
\item \Code{\backsl FADiagram} advances to the next diagram.
\end{itemize}
Since \Code{feynarts.sty} emits direct PostScript primitives, the 
interpretation of which is non-standard across PostScript renderers, it 
is guaranteed to work only with \Code{dvips}.

\subsection{Geometry}

\begin{figure}
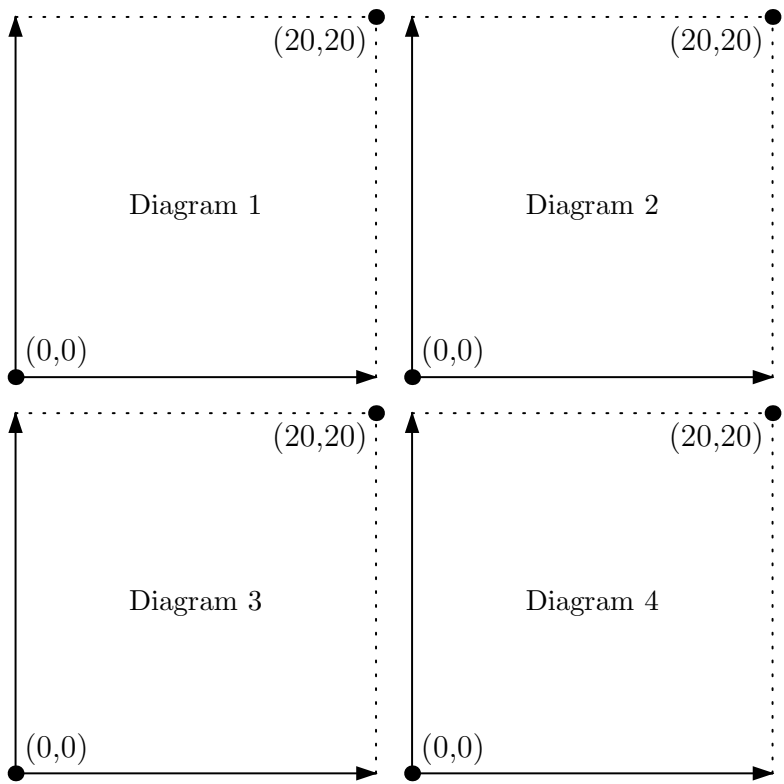

\begin{center}
\begin{feynartspicture}(300,345)(2,2.3)
\FALabel(22,47)[]{\large Title}

\FADiagram{\raise 65bp\hbox{Diagram 1}}
\FAProp(0,0)(20,0)(0,){/Straight}{0}
\FAProp(19,0)(20,0)(0,){/Straight}{1}
\FAProp(0,0)(0,20)(0,){/Straight}{0}
\FAProp(0,19)(0,20)(0,){/Straight}{1}
\FAProp(20,0)(20,20)(0,){/GhostDash}{0}
\FAProp(0,20)(20,20)(0,){/GhostDash}{0}
\FAVert(0,0){0}
\FALabel(.5,.5)[lb]{(0,0)}
\FAVert(20,20){0}
\FALabel(19.5,19.5)[rt]{(20,20)}

\FADiagram{\raise 65bp\hbox{Diagram 2}}
\FAProp(0,0)(20,0)(0,){/Straight}{0}
\FAProp(19,0)(20,0)(0,){/Straight}{1}
\FAProp(0,0)(0,20)(0,){/Straight}{0}
\FAProp(0,19)(0,20)(0,){/Straight}{1}
\FAProp(20,0)(20,20)(0,){/GhostDash}{0}
\FAProp(0,20)(20,20)(0,){/GhostDash}{0}
\FAVert(0,0){0}
\FALabel(.5,.5)[lb]{(0,0)}
\FAVert(20,20){0}
\FALabel(19.5,19.5)[rt]{(20,20)}

\FADiagram{\raise 65bp\hbox{Diagram 3}}
\FAProp(0,0)(20,0)(0,){/Straight}{0}
\FAProp(19,0)(20,0)(0,){/Straight}{1}
\FAProp(0,0)(0,20)(0,){/Straight}{0}
\FAProp(0,19)(0,20)(0,){/Straight}{1}
\FAProp(20,0)(20,20)(0,){/GhostDash}{0}
\FAProp(0,20)(20,20)(0,){/GhostDash}{0}
\FAVert(0,0){0}
\FALabel(.5,.5)[lb]{(0,0)}
\FAVert(20,20){0}
\FALabel(19.5,19.5)[rt]{(20,20)}

\FADiagram{\raise 65bp\hbox{Diagram 4}}
\FAProp(0,0)(20,0)(0,){/Straight}{0}
\FAProp(19,0)(20,0)(0,){/Straight}{1}
\FAProp(0,0)(0,20)(0,){/Straight}{0}
\FAProp(0,19)(0,20)(0,){/Straight}{1}
\FAProp(20,0)(20,20)(0,){/GhostDash}{0}
\FAProp(0,20)(20,20)(0,){/GhostDash}{0}
\FAVert(0,0){0}
\FALabel(.5,.5)[lb]{(0,0)}
\FAVert(20,20){0}
\FALabel(19.5,19.5)[rt]{(20,20)}
\end{feynartspicture}
\end{center}
\caption{\label{fig:geometry}Geometry of a $2\times 2$
\Code{feynartspicture} sheet.}
\end{figure}

A single Feynman diagram is always drawn on a $20\times 20$ canvas. 
Several such canvasses are combined into a rectangular sheet which can
optionally carry a title.  See Fig.~\ref{fig:geometry} for illustration. 
Such a sheet of Feynman diagrams is enclosed in a \Code{feynartspicture}
environment in \LaTeX:
\begin{alltt}
   \backsl{}begin\Brac{feynartspicture}(\(s\sb{x}\),\(s\sb{y}\))(\(n\sb{x}\),\(n\sb{y}\))
   ...
   \backsl{}end\Brac{feynartspicture}
\end{alltt}
This sheet has a size of $s_x\times s_y$ (in units of \LaTeX's
\Code{\backsl unitlength}) with room for $n_x\times n_y$ Feynman
diagrams.  $n_y$ need not be an integer and the extra space implied
by the fractional part is allocated at the top for the sheet label.

Note that it is not possible to distort the aspect ratio of a 
Feynman diagram.  If the ratio $n_x/\lfloor n_y\rfloor$ is chosen 
different from the ratio $s_x/s_y$, the sheet will fit the smaller 
dimension exactly and be centered in the larger dimension.

Inside the \Code{feynartspicture}, the macro
\begin{alltt}
   \backsl{}FADiagram\Brac{\(\Var{dtitle}\)}
\end{alltt}
advances to the next diagram, which has the title \Var{dtitle}.  The 
size of \Var{dtitle} can be changed by redefining \Code{\backsl 
FADiagramLabelSize} with one of the usual \LaTeX\ font-size specifiers,
\eg
\begin{verbatim}
   \renewcommand\FADiagramLabelSize{\scriptsize}
\end{verbatim}
The default size is \Code{\backsl small}.

\subsection{Propagators}

All propagators are circular arcs in the FeynArts style.  This includes
conceptually the straight line as the infinite-radius limit. 
Propagators furthermore come in two variants: tadpole propagators, where
the initial and final vertex coincide, and `ordinary' propagators with
distinct initial and final vertex.  This distinction is necessary
because the information that has to be stored is different for the two
cases.  The arguments of the \Code{\backsl FAProp} macro and their
geometrical meaning are shown in Fig.~\ref{fig:prop} for both variants.

\begin{figure}
\begin{center}
\begin{picture}(440,160)(-50,10)
\SetWidth{1.5}
%
\CArc(50,100)(40,0,360)
\Vertex(50,60){2}
\Text(50,55)[t]{$(f_x,f_y)$}
\Vertex(50,100){2}
\Text(50,98)[t]{$(c_x,c_y)$}
\Text(50,20)[]{\Code{\backsl FAProp($f_x$,$f_y$)($f_x$,$f_y$)($c_x$,$c_y$)\Brac{$g$}\Brac{$a$}}}
\SetOffset(70,0)
\SetWidth{.5}
\SetColor{Blue}
\Line(156.031,84.202)(211.698,116.342)
\Line(211.698,116.342)(200,136.603)
\SetColor{Black}
\DashLine(250,50)(267.365,148.481){4}
\DashLine(250,50)(156.031,84.202){4}
\DashLine(211.698,116.342)(267.365,148.481){4}
\SetWidth{1.5}
\CArc(250,50)(100,80,160)
\Vertex(267.365,148.481){2}
\Vertex(156.031,84.202){2}
\Vertex(250,50){2}
\Text(271,150)[lb]{$(t_x,t_y)$}
\Text(154,82)[rt]{$(f_x,f_y)$}
\Text(185,98)[t]{$d$}
\Text(210,130)[l]{$h$}
\Text(230,100)[]{$\kappa = \dfrac hd$}
\Text(220,20)[]{\Code{\backsl FAProp($f_x$,$f_y$)($t_x$,$t_y$)($\kappa$,)\Brac{$g$}\Brac{$a$}}}
\end{picture}
\end{center}
The latter two arguments, $g$ and $a$, respectively determine line and
arrow style:
\begin{center}
\vspace*{-5ex}
\begin{feynartspicture}(360,180)(2,1)
\FADiagram{}
\FAProp(1,16)(10,16)(0,){/Straight}{0}
\FALabel(11,16)[l]{$g$ = /Straight}
\FAProp(1,13)(10,13)(0,){/ScalarDash}{0}
\FALabel(11,13)[l]{$g$ = /ScalarDash}
\FAProp(1,10)(10,10)(0,){/GhostDash}{0}
\FALabel(11,10)[l]{$g$ = /GhostDash}
\FAProp(1,7)(10,7)(0,){/Sine}{0}
\FALabel(11,7)[l]{$g$ = /Sine}
\FAProp(10,4)(1,4)(0,){/Cycles}{0}
\FALabel(11,4)[l]{$g$ = /Cycles}
\FADiagram{}
\FAProp(3,16)(12,16)(0,){/Straight}{0}
\FALabel(13,16)[l]{$a = 0$}
\FAProp(3,13)(12,13)(0,){/Straight}{1}
\FALabel(13,13)[l]{$a = 1$}
\FAProp(3,10)(12,10)(0,){/Straight}{-1}
\FALabel(13,10)[l]{$a = -1$}
\end{feynartspicture}
\vspace*{-5ex}
\end{center}
Note the slash (/) in the line-style directive which is necessary because
the directive is directly handed to the PostScript interpreter.
\caption{\label{fig:prop}The geometrical layout of a propagator and the
corresponding arguments of the \Code{\backsl FAProp} macro.
Left: tadpole-type propagators (coincident initial and final vertex).
Right: `ordinary' propagators (non-coincident initial and final vertex).}
\end{figure}

\subsection{Vertices}

Vertices mark the points where propagators join.  Each propagator has a 
counter-term order associated with it.  
\begin{center}
\vspace*{-10ex}
\hspace*{20ex}\begin{feynartspicture}(200,200)(1,1)
\FADiagram{}
\FALabel(-1,10)[r]{\Code{\backsl FAVert($x$,$y$)\Brac{$o$}}\qquad $o =~~\cdots$}
\FAVert(1,13){-3}
\FALabel(1,10)[]{$-3$~}
\FAVert(4,13){-2}
\FALabel(4,10)[]{$-2$~}
\FAVert(7,13){-1}
\FALabel(7,10)[]{$-1$~}
\FAVert(10,13){0}
\FALabel(10,10)[]{$0$}
\FAVert(13,13){1}
\FALabel(13,10)[]{$1$}
\FAVert(16,13){2}
\FALabel(16,10)[]{$2$}
\FAVert(19,13){3}
\FALabel(19,10)[]{$3$}
\FALabel(21,10)[l]{$\cdots$}
\end{feynartspicture}
\vspace*{-15ex}
\end{center}

\subsection{Labels}

Labels are usually associated with propagators, but can in principle be
set anywhere.  They have a pair of coordinates and an alignment, given
in the usual \TeX\ manner, \ie a code of up to two letters for vertical
and horizontal alignment: $\{\Code{t} = \text{top}, \text{(empty)} =
\text{center}, \Code{b} = \text{bottom}\}\otimes \{\Code{l} =
\text{left}, \text{(empty)} = \text{center}, \Code{r} = \text{right}\}$,
\eg \Code{[t]} or \Code{[rb]}.  The alignment makes it possible to
change the label's text, in particular its width, without having to
reposition the coordinates.
$$
\Code{\backsl FALabel(\Var{x},\Var{y})[\Var{align}]\Brac{\Var{text}}}
$$

\section{Summary}

FeynEdit is a Java program for editing Feynman diagrams.  It uses the
\LaTeX\ representation of FeynArts for input and output.  Diagrams are 
entered into and retrieved from the editor through cut-and-paste with 
the mouse.  This makes it unnecessary to first save the diagrams one 
wants to edit in a separate file.

The editor does not show details such as line styles and the actual
labels.  This is currently done for performance reasons.  With the
FeynArts \LaTeX\ format, it is not difficult to change these elements
later, however.

The package is open source and is licensed under the LGPL.  It can be 
downloaded from \Code{http://www.feynarts.de} and runs on all platforms 
with a Java interpreter.

\section*{Acknowledgements}

We thank M.~Schmaus for significant help with the Java programming.

\end{document}